\documentclass{ws-ijmpcs}

\usepackage{wrapfig}

\def\trimmarks{}

\graphicspath{{figures/}}

\begin{document}

\markboth{L. Bland}
{Measurement of forward jets at RHIC}

%
\catchline{}{}{}{}{}
%

\title{MEASUREMENT OF FORWARD JETS AT RHIC}

\author{L.~C.~BLAND, for the A$_N$DY collaboration}

\address{Department of Physics, Brookhaven National Laboratory, Upton, NY  USA}

\maketitle

\begin{history}
\end{history}

\begin{abstract}
We present first measurements of forward jet production from $p^\uparrow+p$ collisions 
at $\sqrt{s}=500$ GeV, including transverse single spin asymmetries.  These asymmetries are expected to
be sensitive to spin-correlated transverse momentum in the initial state,
which is particularly interesting because it is related to orbital angular
momentum in the proton.  
\end{abstract}


\ccode{PACS numbers: 12.38.Qk, 13.87-a, 13.88+e}

\section{Introduction}	
The current view is that fast-moving protons consist of collinear
quarks, antiquarks and gluons.  Parton distribution functions (PDF),
at leading order, give the probability to find the parton carrying a
fraction $x$ of the proton momentum.  The PDF are proven to be
universal functions in hard scattering processes based on
factorization theorems.  This picture is appealing and believed to be
complete.  Spin-dependent PDF and their collinear extensions through
generalized parton distributions can be extracted from hard
scattering processes that include polarization degrees of freedom, and
their moments can be used to understand how the proton gets its spin
from its constituents.

There are several puzzles that arise when considering spin.  First,
the quark spins alone cannot account for the spin of the proton (see
Ref.~[\refcite{DSSV}], and references therein).  There must be
contributions from gluon polarization or from orbital angular momentum
(OAM).  Second, there are strikingly large analyzing powers ($A_N$)
measured for $p^{\uparrow}+p\rightarrow\pi+X$ over a broad range of
$\sqrt{s}$ (see Ref.~[\refcite{STARpi0}], and references therein).
Collinear perturbative QCD (pQCD) at leading twist cannot account for
such large spin effects.  Extensions to the theory to include
transverse momentum dependence (TMD) correlated with spin degrees of
freedom were introduced to explain these large spin effects.
Spin-correlated TMD in the initial state [\refcite{Si90}] has been
associated with partonic OAM, albeit in a model dependent way
[\refcite{Ba11}].  Inclusive pion production cannot distinguish
initial-state and final-state spin-correlated TMD [\refcite{Co93}].
This can be distinguished for processes that explicitly include two
scales, such as semi-inclusive deep inelastic scattering (SIDIS)
[\refcite{SIDIS}], or by polarized proton collisions that produce
jets, direct photons or Drell-Yan (DY) dilepton pairs.

Spin physics has seen great recent progress to address these puzzles,
including evidence of non-zero gluon polarization
[\refcite{DSSV2,STAR}], and recent suggestions of how gluon
polarization can be computed in lattice QCD [\refcite{Ji13}].  As
well, we expect to see first measurements of spin observables for
low-mass virtual photons produced by the DY process [\refcite{Qu11}].
An attempt to develop forward instrumentation for spin-dependent DY at
RHIC has led to major plans for the future ({\it e.g.},
Ref. [\refcite{sPHENIX}]).  Tests associated with a proposed
spin-dependent DY measurement at RHIC have led to first measurement of
transverse single spin asymmetries for forward jet production
[\refcite{paper}].  This contribution provides some details beyond the
initial reports of the forward jet measurements.

\begin{wrapfigure}[15]{r}{0.5\textwidth}
\begin{center}
\includegraphics[clip,width=0.48\textwidth]{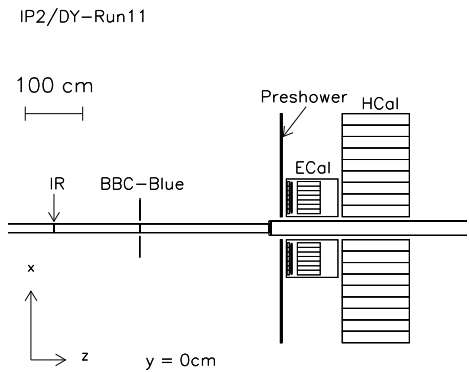}
\caption{A top view from the GEANT model of the 2011 apparatus used
  for the forward jet production measurements.}
\label{apparatus}
\end{center}
\end{wrapfigure}

\section{Experimental Apparatus}

The primary elements of the detector apparatus used for the forward
jet production measurements were left/right symmetric modular hadron
calorimeters (HCal).  Each HCal was a 9-column $\times$ 12-row matrix
of cells.  The cells were originally constructed for AGS-E864
[\refcite{Ar98}], and are 117-cm long lead bars with an embedded
$47\times47$ matrix of 117-cm long scintillating fibers.
Scintillation light is directed onto a single photomultiplier tube
(PMT) from each cell.  Fig. \ref{apparatus} shows a top view of the
apparatus used to view colliding beams in the 2011 run.  Each HCal
module spans $2.5\le\eta\le4.0$ and $|\phi-\phi_{off}|\le0.5$, where
the azimuthal extent refers to the fiducial acceptance used after jet
finding, described below, and $\phi_{off}=0(\pi)$ when the jet is to
the left (right) of the incoming polarized beam.

The two beam-beam counter (BBC) annuli [\refcite{Bi01}] are used to
define a minimum-bias trigger for the events and determine for each
event the $z$ component of the vertex from the distribution of
vertices created by the colliding beam diamond.  The BBC are
scintillator annuli that are mirror symmetric in $z$, where each
annulus spans the azimuth in the $x-y$ plane with 16 trapezoidal
scintillators.  The BBC annuli span the range $2.5 < | \eta | < 3.7$
for collisions at the interaction point.

\begin{figure}[htpb!]
\centering
\begin{tabular}{cc}
\epsfig{file=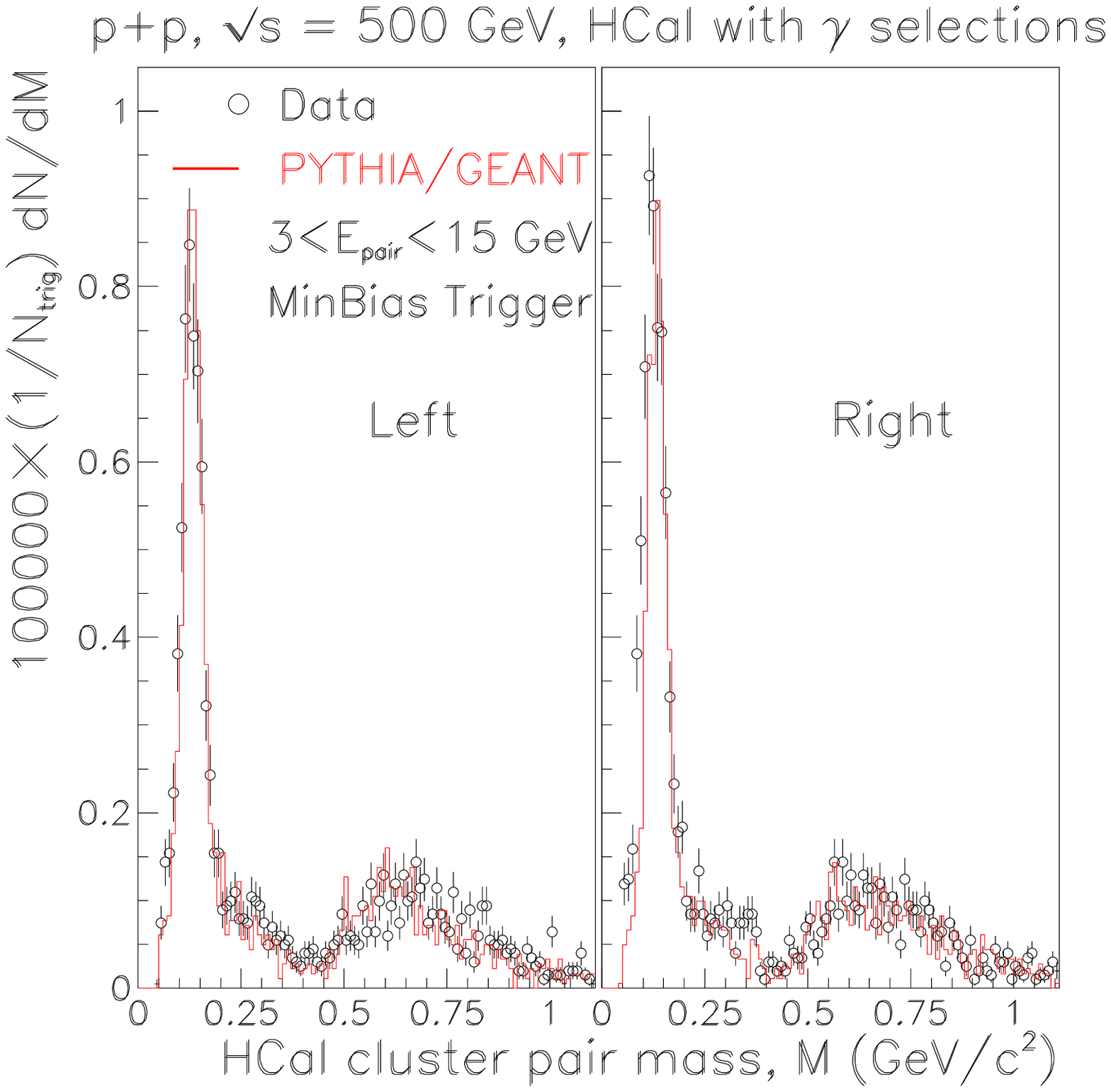,width=0.35\linewidth,clip=}
\hspace{5mm}
\epsfig{file=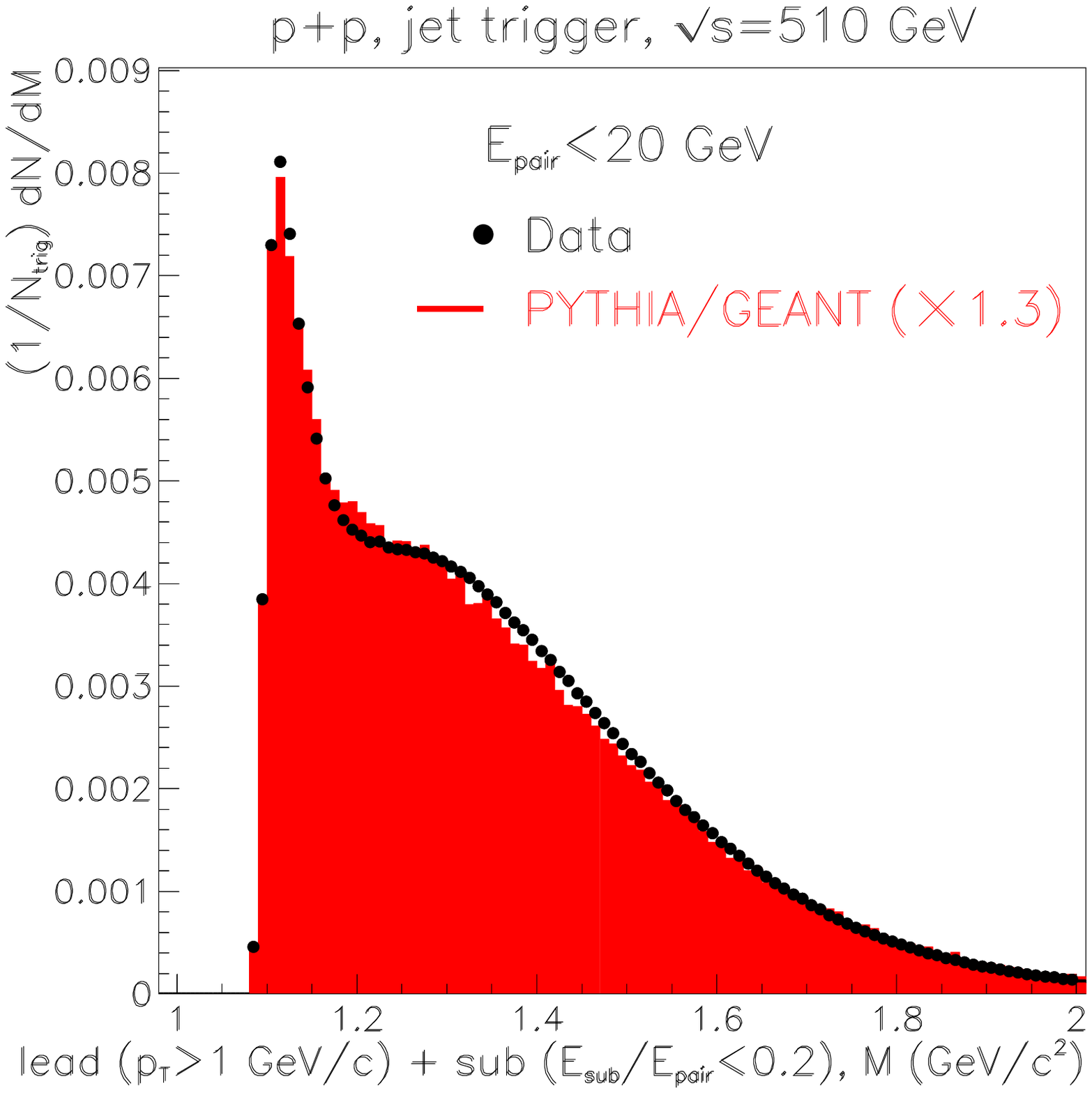,width=0.35\linewidth,clip=}
\end{tabular}
\caption{Cluster pair mass distributions for (left) single-tower
  clusters that are primarily photons, showing
  $\pi^0\rightarrow\gamma\gamma$ and (right) clusters that are not
  tagged as photon-like, where the leading cluster is assigned the
  mass of the proton and the other cluster is assigned the mass of the
  charged pion.  Simulations show this is consistent with
  $\Lambda\rightarrow p\pi^-$ (and conjugate).}
\label{PairMass}
\end{figure}

\section{Calibrations}

Calibrations involve three steps.  The first step is to determine the
energy scale of the HCal.  The second step is to determine
the degree to which the HCal response to incident electromagnetic
(EM) particles ($\gamma,e^\pm$) differs from the response to incident hadrons.
Finally, the HCal responses to different particles are averaged over for
incident jets, so checks of the jet energy scale are important.

Cosmic-ray muons were used prior to the 2011 RHIC run to adjust the
relative gains of the individual PMT for the two HCal modules.  The
absolute energy scale of each HCal module was determined by
reconstruction of $\pi^0\rightarrow\gamma\gamma$ produced by the
colliding beams, by the following procedure.  The collision vertex
was measured for each event by the time difference of the earliest
responding detectors for the two BBC annuli.  Clusters were
reconstructed from the HCal, resulting in cluster energy and
energy-weighted average ($x,y$) positions.  The clusters were assumed
to originate from the event vertex and further assumed to be created
by particles with zero rest mass.  Thus, the event vertex and the
cluster ($x,y,E$) values are sufficient to define the components of a
four momentum.  A simple way to preferentially select clusters from
incident $\gamma$ and $e^\pm$ over incident hadrons is to require the
clusters to consist of only one or two towers, because EM showers have
significantly less transverse extent than do hadronic showers.  Pairs
of these clusters are then used to compute invariant mass for both
data and full simulation ({\it i.e.}, from particles generated by
PYTHIA 6.222 [\refcite{PYTHIA6222}] that are then run through GEANT).
The comparison of absolutely normalized mass distributions
[\refcite{Pe11}] is shown in the left panel of Fig. \ref{PairMass}.

The HCal has an excellent response to incident photons because of its
construction.  Photon test beams used to test similar spaghetti
calorimeters have measured $\sigma_E / E \approx 0.05/\sqrt E$
[\refcite{Le08}], where $\sigma_E$ is the calorimeter resolution in
response to incident energy $E$.  The mass resolution in
Fig. \ref{PairMass} is dominated by the measurement of the diphoton
opening angle, because of the (10 cm)$^2$ cell size of HCal.  As will
be described below, the energy resolution of HCal to photons,
electrons and positrons can be estimated from the reconstruction of
peaks at large mass, since the opening angle resolution is no longer
the limiting factor.

The HCal response is proportional to the energy of incident hadrons,
by design.  Compensation corresponds to the difference in the response
of HCal to different particle types.  Studies of compensation of HCal
have been carried out in simulation, and the expectation is $\sim$20\%
differences in the response from different particles, because the peak
in their shower occurs at different depth in the calorimeter.  Photons
shower close to the entrance of the calorimeter.  The longitudinal
shower profile for charged pions is peaked at $\sim$30 cm depth in the
calorimeter.  Compensation can be checked in the data by
reconstruction of a particle that decays to hadrons, such as
$\Lambda\rightarrow p\pi^-$ and its anti-particle conjugate, since
charge sign is not measured. Given that neutral pions are prolifically
produced, backgrounds for hadronic-like cluster pairs are reduced by
rejecting photon-like clusters.  A crucial step is to make rest-mass
assignments to the clusters.  The leading cluster is assigned the mass
of the proton and the other cluster is assigned the mass of the
charged pion to convert the measured positions and energies of the
clusters into four momenta.  The pair mass distribution shown in the
right panel of Fig.~\ref{PairMass} results.  The centroid of the peak
is insensitive to knowledge of the displaced vertex for the
$\Lambda\rightarrow p\pi^-$ decay (the decay vertex cannot be
reconstructed with the apparatus), as confirmed in full simulations.
The width of the mass peak is weakly sensitive to the HCal resolution,
as demonstrated by adding to the GEANT energy depositions in HCal
towers a Gaussian-distributed smearing.  The HCal tower energy used
for reconstructions is $E=E_{sim}+G\sigma_E$, where
$\sigma_E=b\sqrt{E}$, $G$ is drawn from a Gaussian distribution of
zero mean value and unit $\sigma$, and $b\approx0.34$ for hadronic
showers in HCal [\refcite{Ar98}].

\section{Jet finding and energy scale}

Jets are sprays of primarily mesons and baryons that are localized in
$\eta-\phi$ space.  They are understood to arise from the
fragmentation of hard-scattered quarks and gluons.  They are found by
a pattern recognition algorithm, that identifies energy concentrations
in a circle of radius $R_{jet}$ in $\eta-\phi$ space.  We use the
anti-$k_T$ algorithm [\refcite{Ca08}] for most of our analyses, but
have also considered the mid-point cone algorithm.

Objects to consider in regard to the determination of the jet energy
scale are hard-scattered partons, particle jets and tower jets.  The
hard-scattered partons appear in a conventional computation of
particle production, such as next-to-leading order (NLO) pQCD, or in
event generators such as PYTHIA.  Already this object has
complications because of QCD radiation, computed in PYTHIA via
initial-state (ISS) and final-state (FSS) parton showers.  The FSS
mostly give a finite size to the hard-scattered parton, because the
radiation is distributed about the direction of the parton.  The ISS
can lead to underlying event contributions.  Information about the
hard-scattered parton is inferred from the resulting particle
production that gives rise to the detector response.  Jet-finding
algorithms can be applied to the observable particles that follow all
resonance decays, resulting in particle jets. The energy of this
object can be impacted by particle decay, since decay products can be
distant from the jet in $\eta-\phi$ space.  Finally, the jet-finding
algorithms are applied to the HCal response, presented to the
jet-finder as a table of corrected energy, $\eta$, and $\phi$,
resulting in tower jets.  The $\eta,\phi$ values are determined from
the $x,y$ position of each tower and the distance of the tower from
the collision vertex, assumed to be the source of all particle
production.  The collision vertex is reconstructed for each event from
the time difference of the first arriving particles at each BBC
annulus.

\begin{wrapfigure}[19]{r}{0.5\textwidth}
\begin{center}
\includegraphics[clip,width=0.48\textwidth]{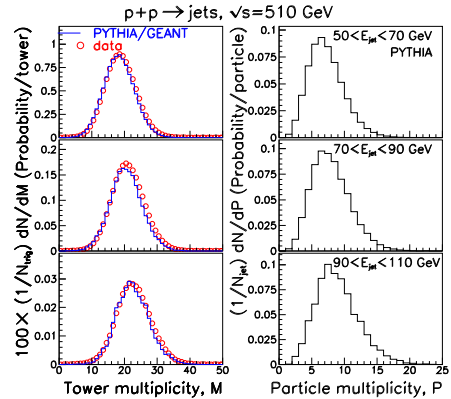}
\vspace*{1pt}
\caption{Multiplicity of HCal towers (left) and stable particles
  (right) from the anti-k$_T$ jet finder applied to data and
  simulation (left) or to stable particles generated by PYTHIA
  6.222 (right).}
\label{TowMult}
\end{center}
\end{wrapfigure}

Checks of the jet-energy scale are most readily made from a simulation
model. The model is validated by comparison of data to full
simulations, that includes a slow simulator applied to accumulated
GEANT hits to produce pseudodata.  The same reconstruction code is run
on data and pseudodata, to facilitate their comparison.  Previous
comparisons of data and simulation have been made for neutral pion
reconstruction [\refcite{Pe11}], summed energy response from HCal
[\refcite{Pe11}], the $p_T$ distribution of reconstructed jets
[\refcite{No12}], and the jet shape [\refcite{No12}].  Another example
of data/simulation comparison is the tower multiplicity in jet-energy
bins (Fig.\ref{TowMult}).  For this figure, the jet trigger used for
the data is emulated for the simulation, and the resulting
multiplicity distributions are scaled by the number of triggers for
both data and simulation.  The full simulation accounts for the shape
and the normalization of the multiplicity of towers associated with
the jets, reconstructed using $R_{jet}$=0.7.  In general, the full
simulation gives a good account of the data.  The multiplicities are
comparable to what is observed in $e^+e^-$ collisions at
$\sqrt{s}\approx10$ GeV [\refcite{CLEO}] and in fixed-target
hadroproduction experiments [\refcite{ftjet}].  The high energies
associated with forward jet production makes their detection possible.

One check of the jet-energy scale is to correlate tower jet energy
with particle jet energy [\refcite{No12}], from the full simulation.
Although it is not commonly done, checks of the jet energy scale can
be obtained directly from the data for some resonance states that
decay to jets, such as $\Upsilon(1S)\rightarrow 3g$.

\begin{wrapfigure}[19]{r}{0.5\textwidth}
\begin{center}
\includegraphics[clip,width=0.48\textwidth]{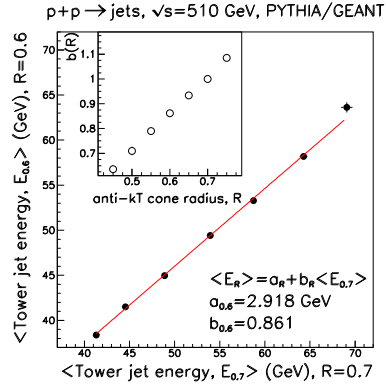}
\vspace*{1pt}
\caption{Dependence of jet energy scale on $R_{jet}$, as determined
  from full PYTHIA/GEANT simulations.  As the jet radius decreases, a
  linear transformation is required to recover the same jet energy scale.
  The inset shows the slope of the linear energy compensation as a
  function of jet size.}
\label{EvsR}
\end{center}
\end{wrapfigure}

In general, for a $p+p$ collision, there can be significant
probability to find multiple jets in an event.  Finite acceptance of
the apparatus can impact multi-jet reconstruction because of the
finite size parameter that enters the jet finder ($R_{jet}$).  The
probability to find multiple jets can be increased by decreasing
$R_{jet}$, as has been commonly done for jet finding at the LHC
[\refcite{CMS12,ATLAS12}].  A natural question that arises when
small-cone jets are reconstructed is whether the energy scale changes
as $R_{jet}$ is decreased.  The correlations between parton energy,
particle jet energy and tower jet energy were used to establish how
the tower-jet energy scale depends on $R_{jet}$.  Results are shown in
Fig.~\ref{EvsR}.  The jet energy scale smoothly varies as the radius
parameter to the jet finder is changed.  The energy scale can be
compensated by a linear transformation, as $R_{jet}$ decreases.  This
variation may result from the low multiplicities for the forward jets.

\begin{figure}[htpb!]
\centering
\begin{tabular}{cc}
\epsfig{file=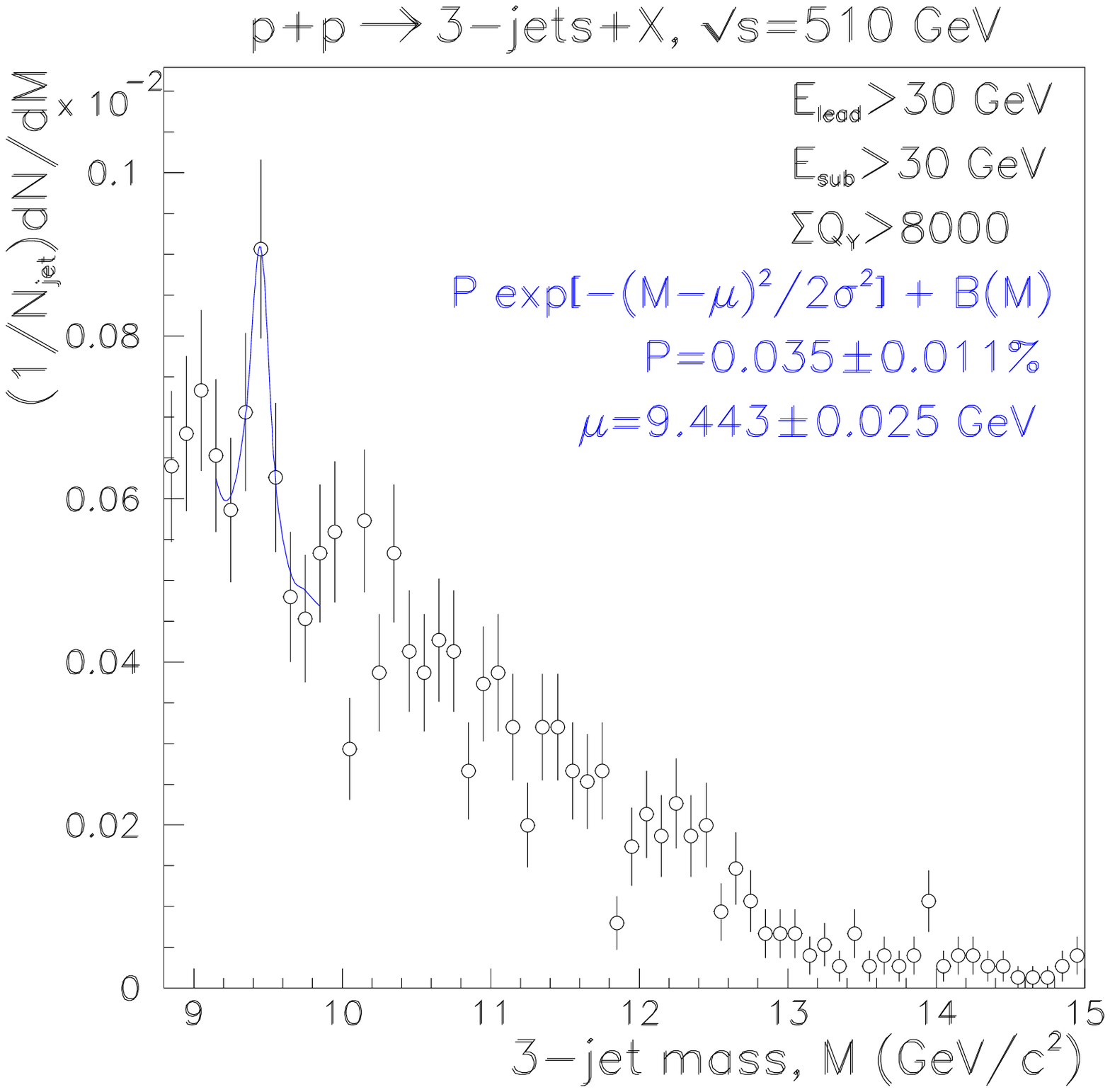,width=0.35\linewidth,clip=}
\hspace{5mm}
\epsfig{file=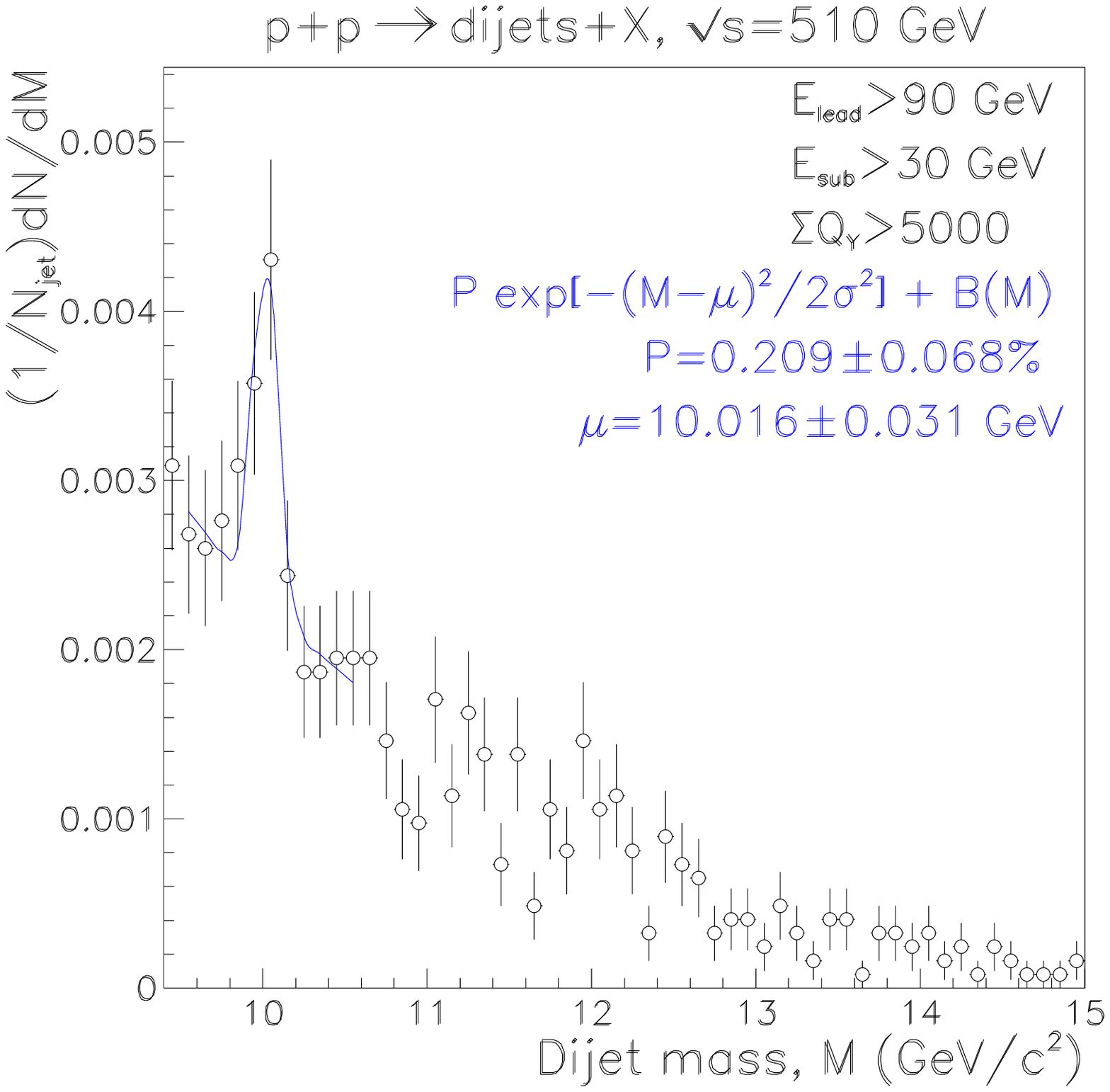,width=0.35\linewidth,clip=}
\end{tabular}
\caption{Mass distributions for 3-jet (left) and 2-jet (right) events,
  with selections on the jet energy and on the multiplicity in the BBC.}
\label{mass}
\end{figure}

Mass distributions for dijet events are discussed below because of
their relevance to the reducible background from conventional QCD
processes to DY production.  Fig.~\ref{mass} shows mass distributions
for a specific event selection for 2- and 3-jet events, where jets are
reconstructed using $R_{jet}$=0.5 and energy compensation from
Fig.~\ref{EvsR}.  The individual jets are assumed to have zero mass,
so the (E,$\eta$,$\phi$) values returned from the jet finder defines a
four vector.  The four-vectors are summed for multi-jet events, and
the mass of the multi-jet system is the magnitude of the four vector.
Events are selected based on jet energies and on the total charge
observed in the BBC annulus that faces the other beam.  A peak is
observed in the 3-jet mass distribution, attributed to
$\Upsilon(1S)\rightarrow 3g$, and a peak is observed in the 2-jet mass
distribution, attributed to $\chi_{2b}(1P)\rightarrow 2g$.

\begin{figure}
\centering
\begin{tabular}{cc}
\epsfig{file=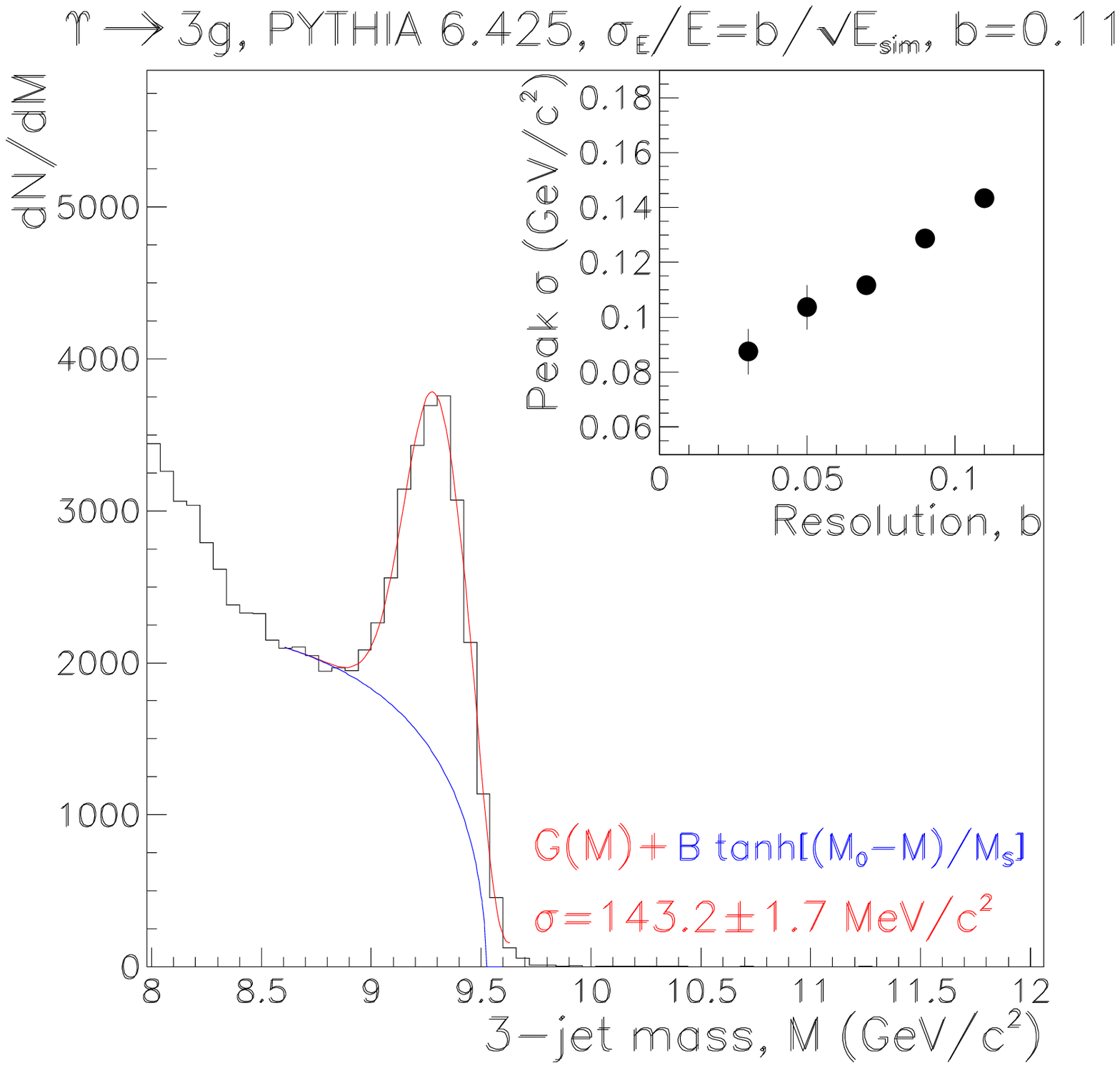,width=0.32\linewidth,clip=}
\hspace{0.5mm}
\epsfig{file=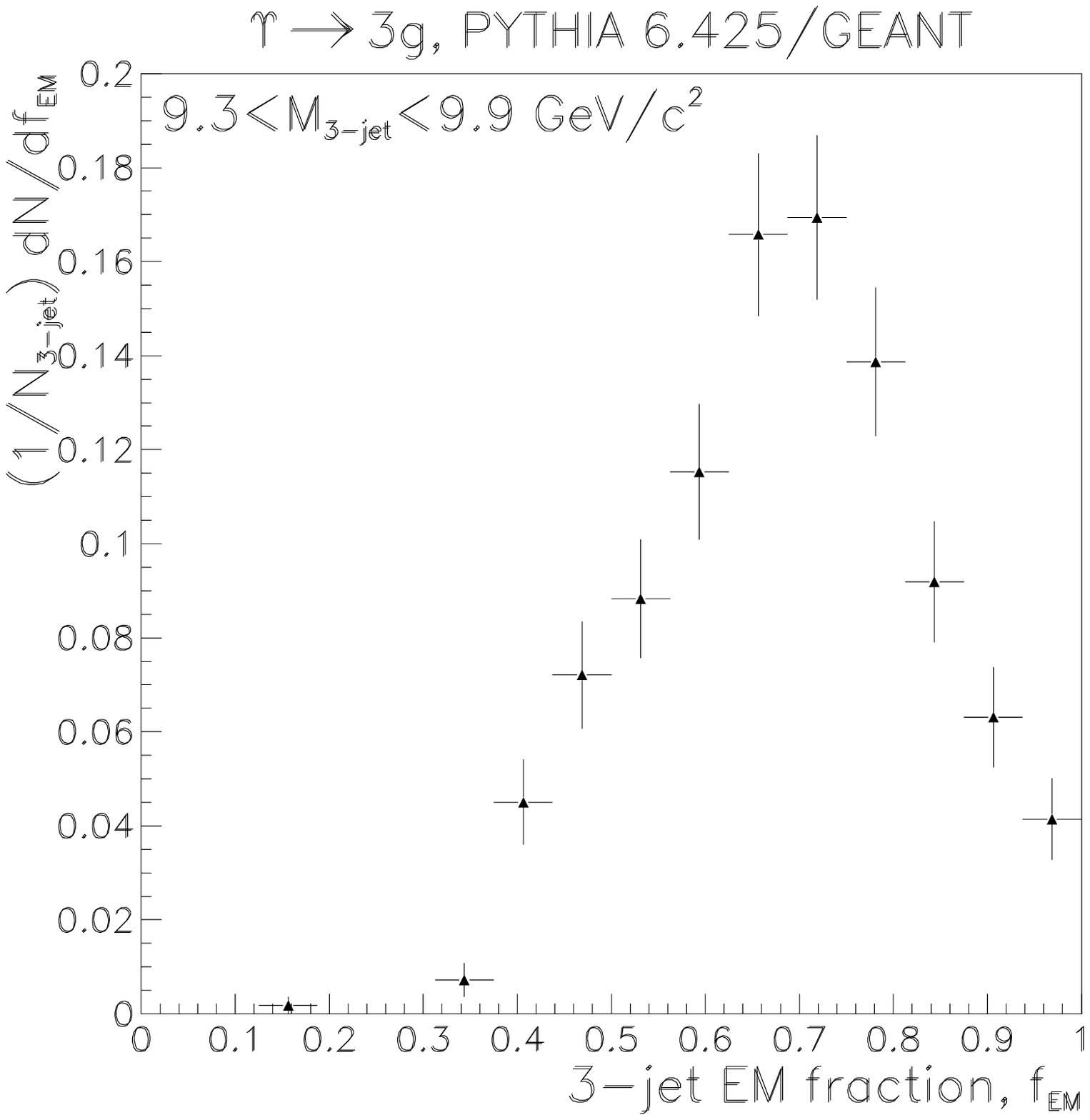,width=0.32\linewidth,clip=}
\hspace{0.5mm}
\epsfig{file=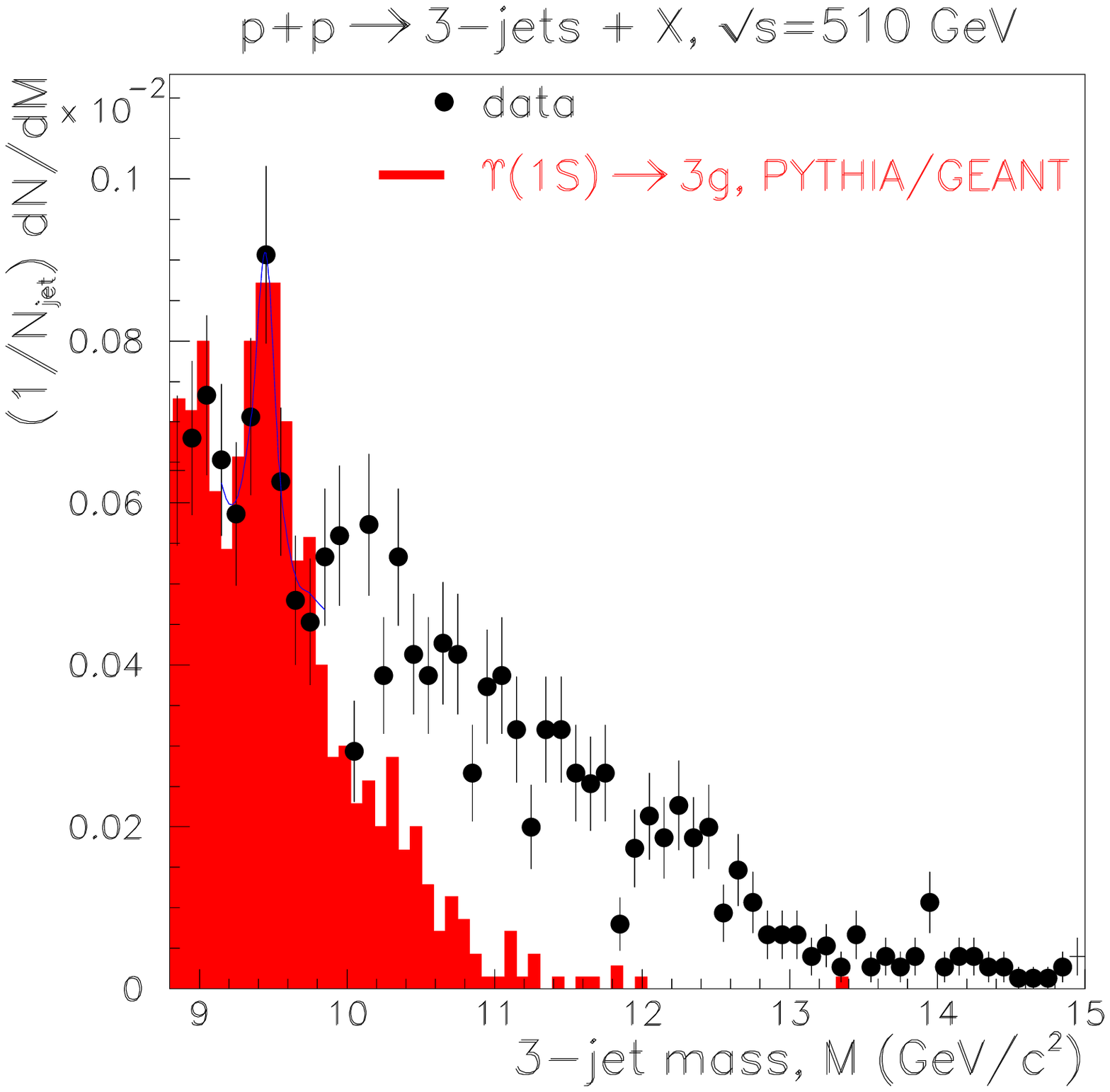,width=0.32\linewidth,clip=}
\end{tabular}
\caption{Results from $\Upsilon(1S)\rightarrow 3g$ simulations (left)
  3-jet mass from the particle jet finder applied to particles the
  3-gluon decay of $\Upsilon(1S)$; (middle) fraction of energy in
  $\gamma, e^\pm$ for full simulations of $\Upsilon(1S)$; and (right)
  comparison of data with reconstruction of full simulation of $\Upsilon(1S)$.}
\label{MultiJetSim}
\end{figure}

A natural question is whether the 3-jet mass peak can be simulated.
Although PYTHIA 6.425 [\refcite{PYTHIA6425}] includes color-singlet
and color-octet matrix elements for bottomonium production, it also
significantly overpredicts the inclusive jet yield, as described
below.  Consequently, we ask a simpler question.  Does
$\Upsilon(1S)\rightarrow 3g$ lead to a peak in the 3-jet mass within
the acceptance of the HCal?  Fig.~\ref{MultiJetSim} shows results from
a particle-jet finder in the left panel.  The primary conclusion here
is that $\Upsilon(1S)\rightarrow 3g$ can lead to a narrow peak in the
3-jet mass distribution, but does require good energy resolution, as
determined by introducing Gaussian smearing of the energy of stable
particles within the HCal acceptance via
$\sigma_E/E=b/\sqrt{E_{sim}}$, before applying the jet finder.

A narrow peak is observed in the 3-jet mass distribution for full
simulations of $\Upsilon(1S)\rightarrow 3g$.  The narrow peak requires
resolution of $\sim 0.05/\sqrt{E}$ for incident $\gamma,e^\pm$, which
is compatible with photon test beam results for similar spaghetti
calorimeters [\refcite{Le08}].  The middle panel of
Fig.~\ref{MultiJetSim} shows the fraction of the 3-jet energy arising
from incident photons, electrons and positrons.  The jets that lead to
the mass peak are dominantly EM fragments.  The right panel of
Fig.~\ref{MultiJetSim} shows the comparison of full simulation of
$\Upsilon(1S)\rightarrow 3g$ to data.  The present simulation does not
attempt to explain the background.

The jet energy scale for inclusive jets is determined from full
simulations of p+p collisions and has been checked for jets that are
dominantly EM fragments of gluons by observation of a
mass peak consistent with $\Upsilon(1S)\rightarrow 3g$.

\section{Results}

\begin{figure}[htpb!]
\centering
\begin{tabular}{cc}
\epsfig{file=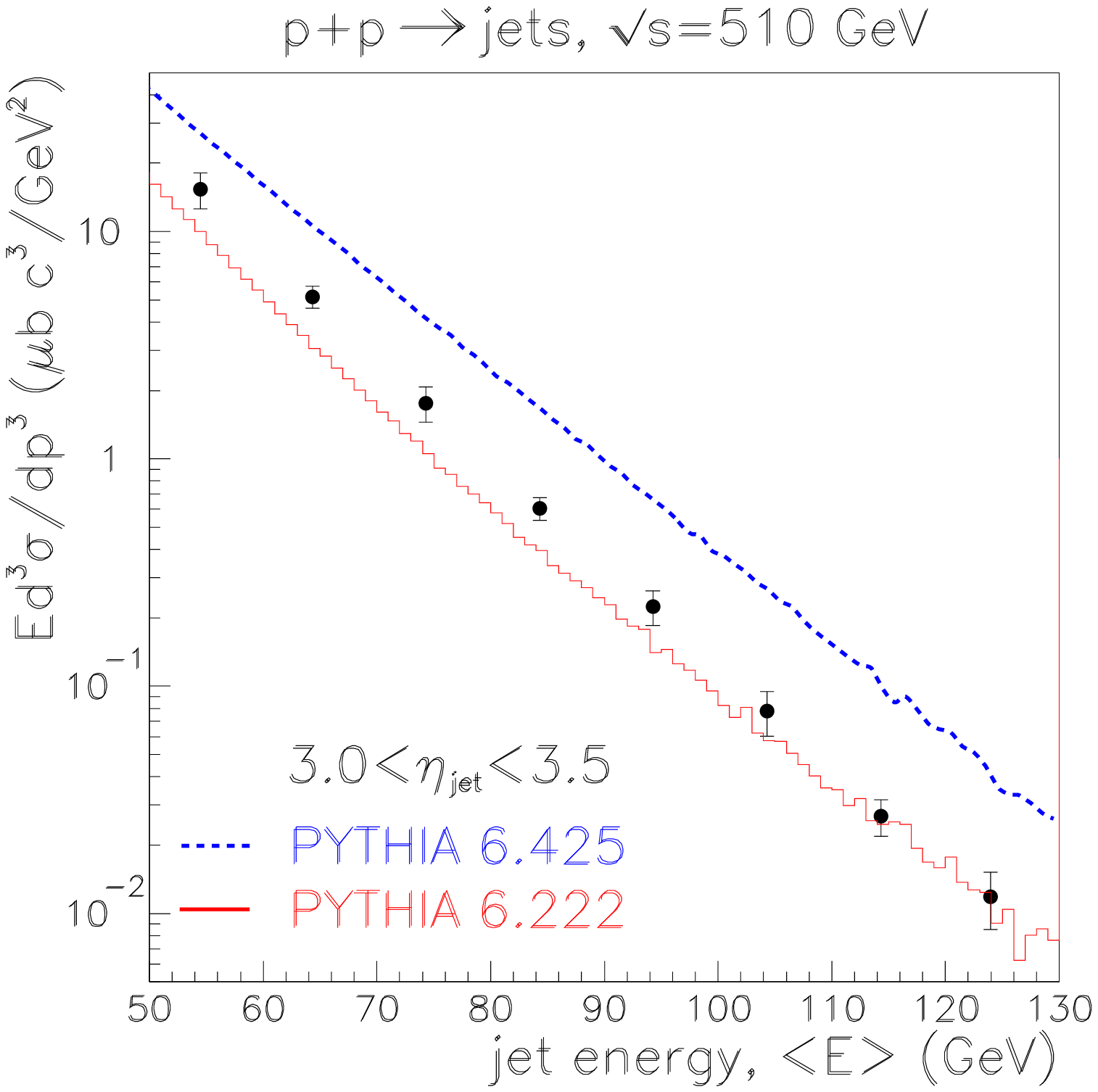,width=0.36\linewidth,clip=}
\hspace{5mm}
\epsfig{file=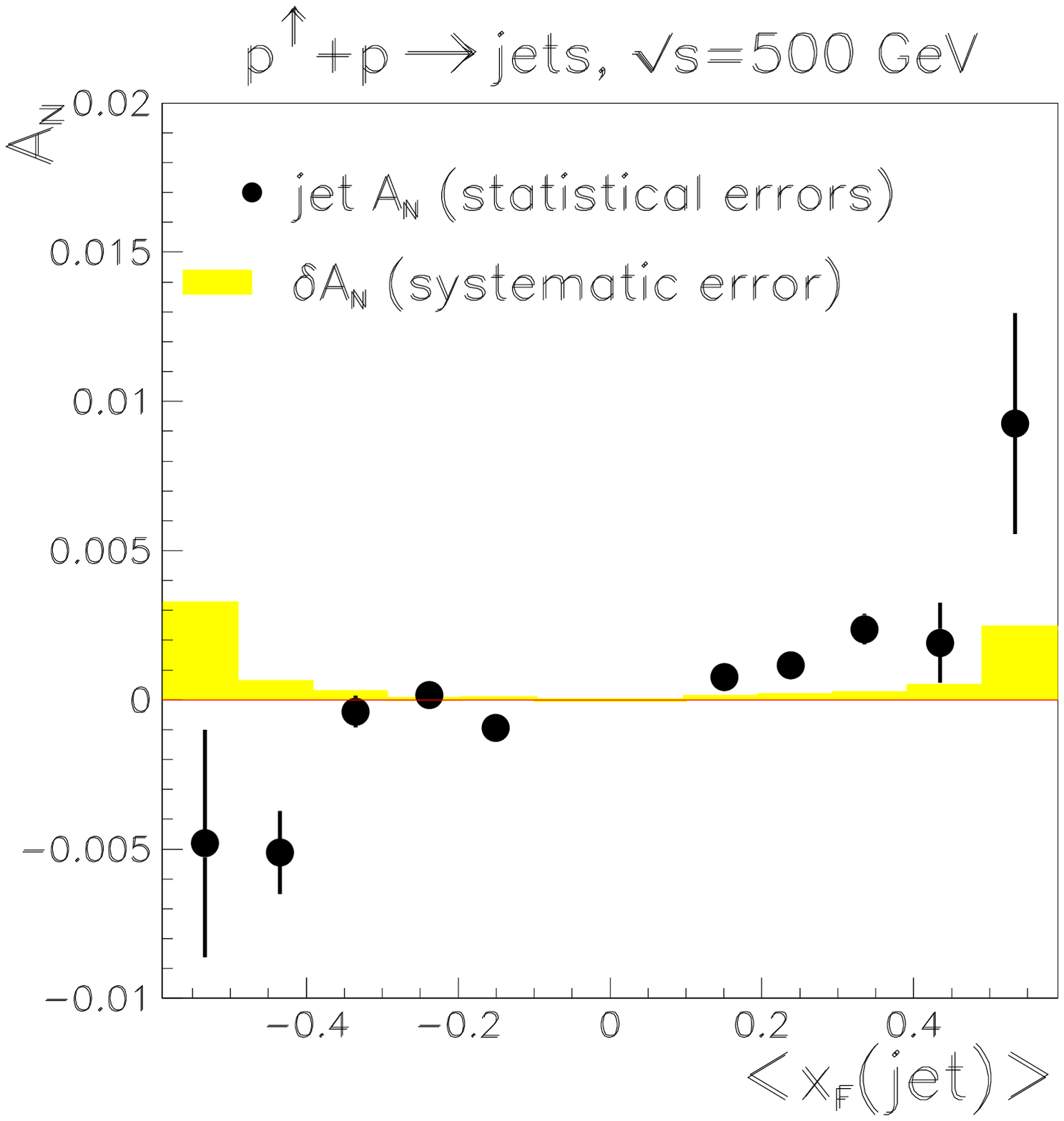,width=0.38\linewidth,clip=}
\end{tabular}
\caption{(left) Cross section for inclusive forward jet production in
  $p+p$ collisions at $\sqrt{s}=510$ GeV.  (right) Analyzing power for
  forward jet production in $p^{\uparrow}+p$ collisions at
  $\sqrt{s}=500$ GeV.}
\label{Results}
\end{figure}

Our measured cross section for inclusive production of forward jets in
$p+p$ collisions at $\sqrt{s}=510$ GeV is shown in the left panel of
Fig.~\ref{Results}, including systematic error estimates previously
described [\refcite{paper}].  The cross sections are compared to
particle jet results from PYTHIA 6.222, that predate tunings to
midrapidity Tevatron measurements for use at the LHC, and PYTHIA 6.425,
that include those tunings.  Similar to forward pion production,
PYTHIA 6.222 gives a better description of forward jet production than
do later versions with underlying event adjustments.  Previously
[\refcite{No12}], we showed the jet cross section was from partonic hard
scattering.  Forward jet $A_N$, previously described [\refcite{paper}]
including the estimates of systematic uncertainties, is shown in the
right panel of Fig.~\ref{Results}. The measured cross section has, so
far, been compared to a pQCD model that assumes the applicability of
factorization for TMD distribution functions [\refcite{An13}].  Good
agreement with the calculation is found.  The jet $A_N$ has been
compared to pQCD model calculations using TMD distribution functions
[\refcite{An13}], but excluding color-charge interactions, and to
NLO, twist-3, collinear pQCD calculations [\refcite{Ga13}].
In general, both calculations give a fair description of the data.
The latter calculation [\refcite{Ga13}] has claimed an indication of the
expected process dependence of spin-correlated $k_T$ in the initial
state.

\begin{wrapfigure}[17]{r}{0.5\textwidth}
\begin{center}
\includegraphics[clip,width=0.48\textwidth]{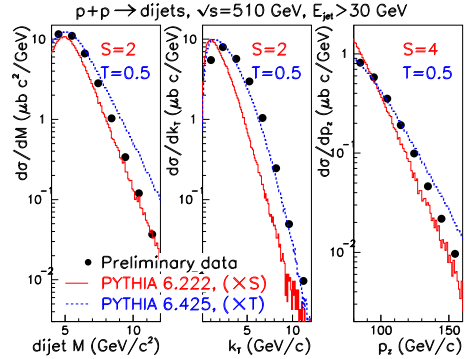}
\vspace*{1pt}
\caption{Dijet cross sections as functions of dijet mass (M),
  dijet momentum imbalance ($k_T$) and dijet longitudinal momentum
  ($p_z$) compared to PYTHIA 6.222 and PYTHIA 6.425}
\label{dijet}
\end{center}
\end{wrapfigure}

Dijet cross sections are of intrinsic interest, because they select
scatterings that primarily involve low-$x$ gluons, and are of
practical interest for any future attempt to measure forward DY
production.  Fig.~\ref{dijet} shows dijet cross sections as functions
of DY kinematic variables, including dijet mass ($M$), dijet momentum
imbalance ($k_T$) and dijet longitudinal momentum ($p_z$).
Corrections to raw yields from jets reconstructed with $R_{jet}=0.7$
follow analogous methods developed for the inclusive jets
[\refcite{paper}], and were checked by verifying that corrected tower
dijet cross sections agreed with input particle jet cross sections for
full simulation.  The dijet cross sections are compared to PYTHIA
6.222 simulations (used for the reducible background estimates for DY
production) and to PYTHIA 6.425.  PYTHIA 6.222 underpredicts the
yields by a factor of two and PYTHIA 6.425 overpredicts the yields by
a factor of two, but can explain the $<k_T>$ for the dijets.

\section{Conclusions}

In conclusion, we have made first measurements of forward jet
production in $p^\uparrow+p$ collisions at $\sqrt{s}=500$ GeV.  Our
measured cross section is consistent with dominant contributions from
partonic hard scattering.  Our measured forward jet $A_N$ is small and
positive, and is compatible with pQCD calculations that fit SIDIS
results for spin-correlated $k_T$ in the initial state, thereby
constraining models that aim to determine partonic OAM.  Dijet results
will be important for low-$x$ physics studies.  Our measured cross
sections determine the reducible background for future forward DY
production.  Observation of $\Upsilon (1S)$ and $\chi_{2b}(1P)$
production through multi-jet final states suggests that the
irreducible background from open heavy flavor production can be
accessed when these bottomonium states are observed through their
dilepton decays.  It remains the case that the most definitive
experiment to test present understanding is a measurement of $A_N$ for
DY production.

\end{document}